\documentclass[11pt]{prop_axis} 

\usepackage[super,comma]{natbib} 
\usepackage{apjfonts} 
\usepackage{graphicx} 
\usepackage{amssymb,amsmath} 
\usepackage{color} 
\usepackage{enumitem} 
\usepackage{wrapfig} 
\usepackage{tikz} 
\usepackage{authblk}

\usepackage{multirow} 
\usepackage{tabularx} 
\usepackage{array} 

\usepackage{hyperref} 
\hypersetup{
    colorlinks,
    linkcolor={blue!80!black},
    citecolor={blue!80!black},
    urlcolor={blue!80!black}
}

\usepackage{sidecap}
\sidecaptionvpos{figure}{c} 

\newcolumntype{L}[1]{>{\raggedright\let\newline\\\arraybackslash\hspace{0pt}}p{#1}}
\newcolumntype{C}[1]{>{\centering\let\newline\\\arraybackslash\hspace{0pt}}p{#1}}
\newcolumntype{R}[1]{>{\raggedleft\let\newline\\\arraybackslash\hspace{0pt}}p{#1}}
\renewcommand\arraystretch{1.3}

\definecolor{headcolor}{rgb}{0.65,0.65,0.65}
\usepackage{fancyhdr}
\renewcommand{\topmargin}{-9mm}

\renewcommand{\textfloatsep}{4mm}

\newcommand{\arcsec}{$^{\prime\prime}$}

\newcommand{\arcmin}{$^{\prime}$}

\newcommand{\bsf}{\sffamily\bfseries}

\definecolor{callout}{rgb}{0.25,0.40,0.85}
\definecolor{synergies}{rgb}{0.20,0.45,0.99}
\definecolor{methods}{rgb}{0.20,0.70,0.45}

\definecolor{calllem}{rgb}{0.20,0.45,0.99}
\definecolor{tabledef}{rgb}{0.95,0.95,0.95}
\definecolor{tablealt}{rgb}{0.77,0.80,1.0}
\definecolor{tablelem}{rgb}{0.80,0.85,1.0}
\definecolor{whitelem}{rgb}{1.0,1.0,1.0}
\definecolor{greenlem}{rgb}{0.7,1.0,0.7}

\newcommand{\golem}{\textcolor{black}{{\em LEM}}}
\def\pluto{{\sc pluto}}

\author[1]{S.\,Orlando}
\author[2,1]{M.\,Miceli}
\author[3]{D.J.\,Patnaude}
\author[3]{P.P.\,Plucinsky}
\author[4,5]{S.-H.\,Lee}
\author[6,7]{C.\,Badenes}
\author[8]{H.-T.\,Janka}
\author[8]{A.\,Wongwathanarat}
\author[3]{J.\,Raymond}
\author[9]{M.\,Sasaki}
\author[8]{E.\,Churazov}
\author[10,8]{I.\,Khabibullin}
\author[1]{F.\,Bocchino}
\author[3]{D.\,Castro}
\author[11]{M.\,Millard}

\affil[1]{INAF - Osservatorio Astronomico di Palermo, Piazza del Parlamento 1, 90134 Palermo, Italy}
\affil[2]{Dip. di Fisica e Chimica, Universit\'a di Palermo, via Archirafi 39, Palermo, Italy}
\affil[3]{Harvard-Smithsonian Center for Astrophysics, MS-3, 60 Garden Street, Cambridge, MA 02138, USA}
\affil[4]{Department of Astronomy, Kyoto University Oiwake-cho, Kitashirakawa, Sakyo-ku, Kyoto 606-8502, Japan}
\affil[5]{Kavli Institute for the Physics and Mathematics of the Universe (WPI), The University of Tokyo, Kashiwa 277-8583, Japan}
\affil[6]{Department of Physics and Astronomy, University of Pittsburgh, 3941 O’Hara Street, Pittsburgh, PA 15260, USA}
\affil[7]{Pittsburgh Particle Physics, Astrophysics and Cosmology Center, University of Pittsburgh, Pittsburgh, PA 15260, USA}
\affil[8]{Max Planck Institute for Astrophysics, Karl-Schwarzschild-Stra\ss e 1, D-85748 Garching, Germany}
\affil[9]{Dr. Karl Remeis Observatory, Erlangen Centre for Astroparticle Physics, Friedrich-Alexander-Universit\"at Erlangen-N\"urnberg, Sternwartstraße 7, 96049 Bamberg, Germany}
\affil[10]{Universitäts-Sternwarte, Ludwig-Maximilians-Universität München, Scheinerstr.1, 81679 München, Germany}
\affil[11]{Department of Physics and Astronomy, University of Iowa, 203 Van Allen Hall, Iowa City, IA 52242-1479, USA}

\begin{document}

\baselineskip=13.2pt
\sloppy
\pagenumbering{roman}
\thispagestyle{empty}

\title{\textcolor{headcolor}{\LARGE\bsf 
Unveiling the Physics of Core-Collapse Supernovae with the Line Emission Mapper: Observing Cassiopeia~A}}
\maketitle
\vspace*{-10mm}

\begin{tikzpicture}[remember picture,overlay]
\node[anchor=north west,yshift=2pt,xshift=2pt]%
    at (current page.north west)
    {\includegraphics[height=20mm]{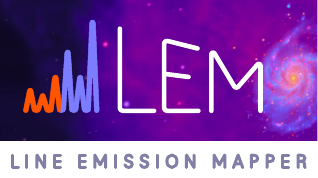}};
\end{tikzpicture}

\section*{SUMMARY}
\label{sec:summary}

Core-collapse supernova remnants (SNRs) display complex morphologies and asymmetries, reflecting anisotropies from the explosion and early interactions with the circumstellar medium (CSM). Spectral analysis of these remnants can provide critical insights into supernova (SN) engine dynamics, the nature of progenitor stars, and the final stages of stellar evolution, including mass-loss mechanisms in the millennia leading up to the SN event.

This white paper evaluates the potential of the Line Emission Mapper (\golem), an advanced X-ray probe concept proposed in response to NASA’s 2023 APEX call, to deliver high-resolution spectra of SNRs. Such capabilities would allow detailed analysis of parent SNe and progenitor stars, currently beyond our possibilities. We employed a state-of-the-art hydrodynamic model that simulates the evolution of a neutrino-driven SN from core-collapse to a 2000-year-old mature remnant. This model successfully replicates the large-scale properties of Cassiopeia A, one of the most studied SNRs in our Galaxy, at an age of about 350 years. Using this model, we synthesized mock {\golem} spectra from different regions of the remnant at different evolutionary phases, accounting for: (i) line shifts and broadening resulting from plasma bulk motion and thermal ion motion, (ii) deviations from ionization equilibrium and electron-proton temperature equilibrium, and (iii) photoelectric absorption by the interstellar medium.

Using standard data analysis tools, we analyzed mock {\golem} spectra as if they were genuine, unveiling the formidable capabilities of {\golem}. We demonstrate that fitting these spectra with phenomenological plasma components enables the accurate recovery of the line-of-sight velocity of the ejecta. This opens avenues for the exploration of the three-dimensional structure of shocked ejecta, akin to methodologies employed in the optical band. Our analysis further indicates {\golem}'s ability in distinguishing between bulk Doppler broadening and thermal broadening of the ion lines. Notably, we can accurately measure ion temperatures near the remnant limb, enabling investigations into ion heating at the shock front and ion cooling in the post-shock flow. This study sheds light on the remarkable potential of {\golem} in advancing our understanding of core-collapse SN dynamics and associated physical processes.

\vfill

\centerline{\em White Paper, August 2024}

\clearpage
\twocolumn


\setcounter{page}{1}
\pagenumbering{arabic}

\section{Introduction}
\label{sec:intro}

Core-collapse supernovae (SNe), marking the endpoint of massive star lifecycles, exert a significant influence on the dynamical and chemical evolution of galaxies \cite{2017hsn..book.....A}. These catastrophic events inject essential ingredients, including mass and heavy elements, into the interstellar medium (ISM), along with substantial energy contributions in the forms of kinetic energy, thermal energy, and cosmic rays (CRs) \cite{2017hsn..book.1981R}. As a result, they drive the chemical enrichment and thermal evolution of the diffuse gas while shaping the structure and dynamics of the ISM. However, despite their fundamental role, many aspects regarding the mechanisms underlying the SN engine and the final phases of stellar progenitor evolution remain poorly understood.

The challenge arises from the spatially unresolved nature of SNe observed in distant galaxies, limiting detailed investigations into the expanding debris and impeding the extraction of critical information concerning progenitors and explosion physics \cite{2017hsn..book.2211M}. These limitations can be addressed by studying nearby young and middle-aged SN remnants (SNRs), namely the outcome of SN explosions. These objects appear as extended sources characterized by a rather complex morphology and a highly non-uniform spatial distribution of ejecta. Part of their features reflect the inner mechanisms of the SN engine, such as those determining the nucleosynthetic yields and the large-scale asymmetries left from the earliest phases of the SN explosion \cite{2015A&A...577A..48W, 2017ApJ...842...13W, 2021A&A...645A..66O}. Other features originate from the interaction of the remnant with an inhomogeneous circumstellar medium (CSM) or may reflect the internal structure of the progenitor star at collapse \cite{2015ApJ...810..168O, 2020ApJ...888..111O, 2020A&A...636A..22O}. 

By conducting spatially resolved spectroscopic analyses of SNRs, one can potentially explore the intricate structure and distribution of chemical elements within their interiors, revealing complex morphologies and highly non-uniform ejecta distributions resulting from the SN explosion and the interaction with the CSM \cite{2012A&ARv..20...49V}. The analysis of high-resolution spectra of SNRs, therefore, holds the promise of providing significant insights into various aspects of astrophysics, including the underlying physical mechanisms driving SN explosions, probing the structure of the CSM, and identifying the nature of the progenitor stellar systems \cite{2012A&ARv..20...49V, 2019NatAs...3..236M}.

The Line Emission Mapper\footnote{WWW: \href{https://lem-observatory.org}{lem-observatory.org} - X / twitter: \href{https://www.twitter.com/LEMXray}{LEMXray} - 
facebook: \href{https://www.facebook.com/LEMXrayProbe}{LEMXrayProbe}.
} ({\golem}) is an innovative X-ray probe concept developed in response to NASA's APEX call for proposals in 2023\cite{2023HEAD...2011018K, Patnaude23}. The instrument consists of a hyperboloid-hyperboloid X-ray optic with a current best estimate effective area of 1500~cm$^{2}$ at 0.5 keV, and a half power diameter of 18\arcsec (see \citet{2023HEAD...2011018K} for details). For comparison, this effective area is expected to be more than an order of magnitude higher than that of XRISM at 1 keV. {\golem}'s focal plane includes an X-ray microcalorimeter with a wide 30\arcmin\ equivalent diameter hexagonal field of view and 15\arcsec\ pixels. In fact, {\golem} will have an angular resolution which is significantly better than that of XRISM Resolve (75\arcsec\ pixels) and this is a great advantage to perform spatially resolved spectroscopy of extended sources as SNRs. Single pixel transition edge sensors (TES) within the inner 5\arcmin\ provide exceptional 1.3~eV resolution, while the outer array consists of 2$\times$2 pixel "hydras" \cite{2020JLTP..199..330S} with a 2.5~eV resolution. The instrument is optimized for the $[0.2,2]$~keV energy band.

Thus, {\golem} has the capacity to efficiently map large diffuse objects such as Galactic SNRs with moderate spatial resolution (five times better than XRISM Resolve) and high spectral resolution (approximately 3.5 times better than XRISM Resolve). Leveraging these capabilities, we can potentially reconstruct the intricate structure of ejected material by measuring the velocities of various elements, such as iron (Fe), titanium (Ti), and silicon (Si), which are crucial for understanding explosion physics. The high-resolution spectra from {\golem} will facilitate an in-depth investigation of key physical processes, such as thermal broadening, charge exchange, and resonant scattering. These observations can also provide detailed insights into non-thermal emission generated by particle acceleration at shock fronts. Furthermore, these measurements enable the evaluation of deviations from ionization equilibrium and electron-proton temperature equilibrium within shocked, swept-up material. Such insights are essential for: 1) probing the physics of SN engines by providing insight into the asymmetries that occurred during the SN explosion, and 2) investigating the endpoint in the evolution of massive stars and the still uncertain physical mechanisms that drive their mass loss. These capabilities of {\golem} are particularly significant given that current grating spectrometers, such as the Chandra HETGS/LETGS and XMM-Newton RGS, encounter difficulties when observing extended sources due to a reduction in their resolving power in these scenarios \cite{2002hrxs.confE..14D}.

In this white paper, we adopt a cutting-edge hydrodynamic model that describes the whole three-dimensional (3D) evolution of a neutrino-driven SN from the core-collapse to its full-fledged remnant \cite{2021A&A...645A..66O}. The model is used to synthesize mock spectra of the SNR at different ages as they would be collected with {\golem}. Notably, the modeled remnant at the age of $\approx 350$~yr matches most of the large-scale properties observed in the galactic SNR Cassiopeia~A (Cas~A). Through the analysis of the synthetic spectra as if they were true spectra, we demonstrate how {\golem} observations will serve as a crucial link between SNRs and their parent SNe and progenitor stars, thereby shedding light on the intricate physics governing SNe and providing crucial insights into the concluding phases of massive star evolution.

The paper is structured as follows: in Section~\ref{sec:model} we summarize the main properties of the adopted SNR model and the synthesis of {\golem} observations; in Section~\ref{sec:analysis} we present the analysis of {\golem} spectra for the remnant at the age of Cas~A ($\approx 350$~years) and at a later stage ($\sim 2000$~years); finally, in Section~\ref{sec:conclusions} we draw our conclusions.
\bigskip

\section{Generation of Mock LEM Spectra}
\label{sec:model}

Cas~A is one of the most studied SNRs in the Galaxy. Due to its proximity, approximately 3.4~kpc away\cite{1995ApJ...440..706R}, and its relative youth of approximately 350 years, Cas~A is a unique laboratory for unraveling the intricate physics of SNe and probing the CSM shaped by the winds of the progenitor star in the millennia leading up to the SN. At all wavelength bands, large-scale asymmetries are evident in the structure and chemical composition of the remnant, likely inherited from the parent SN explosion. Moreover, recent observations\cite{2022ApJ...929...57V, 2024ApJ...965L..27M} and numerical modeling\cite{2022A&A...666A...2O} have provided evidence of the remnant's interaction with an asymmetric dense circumstellar shell, most likely the relic of a massive eruption from the progenitor star that occurred millennia prior to the SN. This scenario was recently supported by JWST observations of Cas~A, which revealed a peculiar extended structure near the center of the remnant, known as the "green monster" \cite{2024ApJ...965L..27M} (see al De Looze et al. 2024, submitted to ApJ). This structure consists of dense, shocked CSM material. One possible interpretation is that the green monster is a relic of the shocked shell.

We expect, therefore, that thermal and non-thermal emission from Cas~A contain crucial information about the dynamics of the explosion, the composition of ejected material, the structure of the CSM, and the mass-loss history of the progenitor star. For these reasons, Cas~A would be an ideal target for {\golem} to gain physical insight about the SN explosion mechanism and the final phases in the evolution of the progenitor star. In this paper, we explore the capabilities of {\golem} in potentially extracting this information by analyzing mock {\golem} spectra of the remnant of a neutrino-driven core-collapse SN that replicates most of the features observed in Cas~A. More specifically, we will use two simulations previously published in the literature: the W15-2-cw-IIb-HD+dec model \cite{2021A&A...645A..66O}, which describes the remnant's expansion through a spherically symmetric wind from the progenitor, and the W15-IIb-sh-HD-1eta-az model \cite{2022A&A...666A...2O}, which details the interaction of the remnant with an asymmetric circumstellar shell.

The adopted models comprehensively describe the 3D evolution of a neutrino-driven SN explosion, spanning from the core-collapse phase to the formation and interaction of its mature remnant with the CSM\cite{2021A&A...645A..66O}. These models integrate the intricate 3D hydrodynamic simulation of a neutrino-driven SN explosion\cite{2017ApJ...842...13W} with hydrodynamic simulations of remnant evolution following the breakout of the shock wave at the stellar surface\cite{2016ApJ...822...22O}. Thus the models describe the development and evolution of the remnant, self-consistently incorporating anisotropies that arise stochastically due to convective overturn in the neutrino-heating layer and activity from the standing accretion shock instability (SASI) occurring during the initial seconds following core bounce. These models are particularly well-suited for our purposes as they capture the emergence of three prominent Fe-rich fingers that naturally, albeit stochastically, form from the earliest stages of the explosion, thus potentially corresponding to the extended shock-heated Fe-rich regions observed in Cas~A.

Both models trace the evolution of explosive nucleosynthesis products generated within the first seconds of the explosion (refer to \citealt{2013A&A...552A.126W, 2015A&A...577A..48W} for details), following their progression until the formation of the fully developed remnant at the age of Cas~A. Furthermore, they extend the evolution to the remnant's age of 2000 years to investigate whether the signatures of the SN explosion remain evident in a more mature SNR and to compare the model with existing, more evolved remnants. The nuclear species under consideration include: protons ($^{1}$H), $^{4}$He, $^{12}$C, $^{16}$O, $^{20}$Ne, $^{24}$Mg, $^{28}$Si, $^{40}$Ca, $^{44}$Ti, $^{56}$Ni, and an additional tracer $^{56}${\em X}, representing Fe-group species synthesized in neutron-rich environments\cite{2015A&A...577A..48W}. Additionally the models include the energy deposition from the dominant radioactive decay chain, wherein $^{56}$Ni decays into $^{56}$Co, and the latter decays into stable $^{56}$Fe. The effects of this energy deposition are relevant for the remnant dynamics and energetics during the first year of its evolution, leading to inflation of initially Ni-rich ejecta\cite{2021A&A...645A..66O, 2021MNRAS.502.3264G}. The simulations were conducted using \pluto, a parallel modular Godunov-type code designed primarily for astrophysical applications and high Mach number flows in multiple spatial dimensions \cite{2012ApJS..198....7M}.

\begin{figure}[!t]
\centering
\includegraphics[width=0.47\textwidth, trim={0cm 0cm 0cm 0cm},clip]{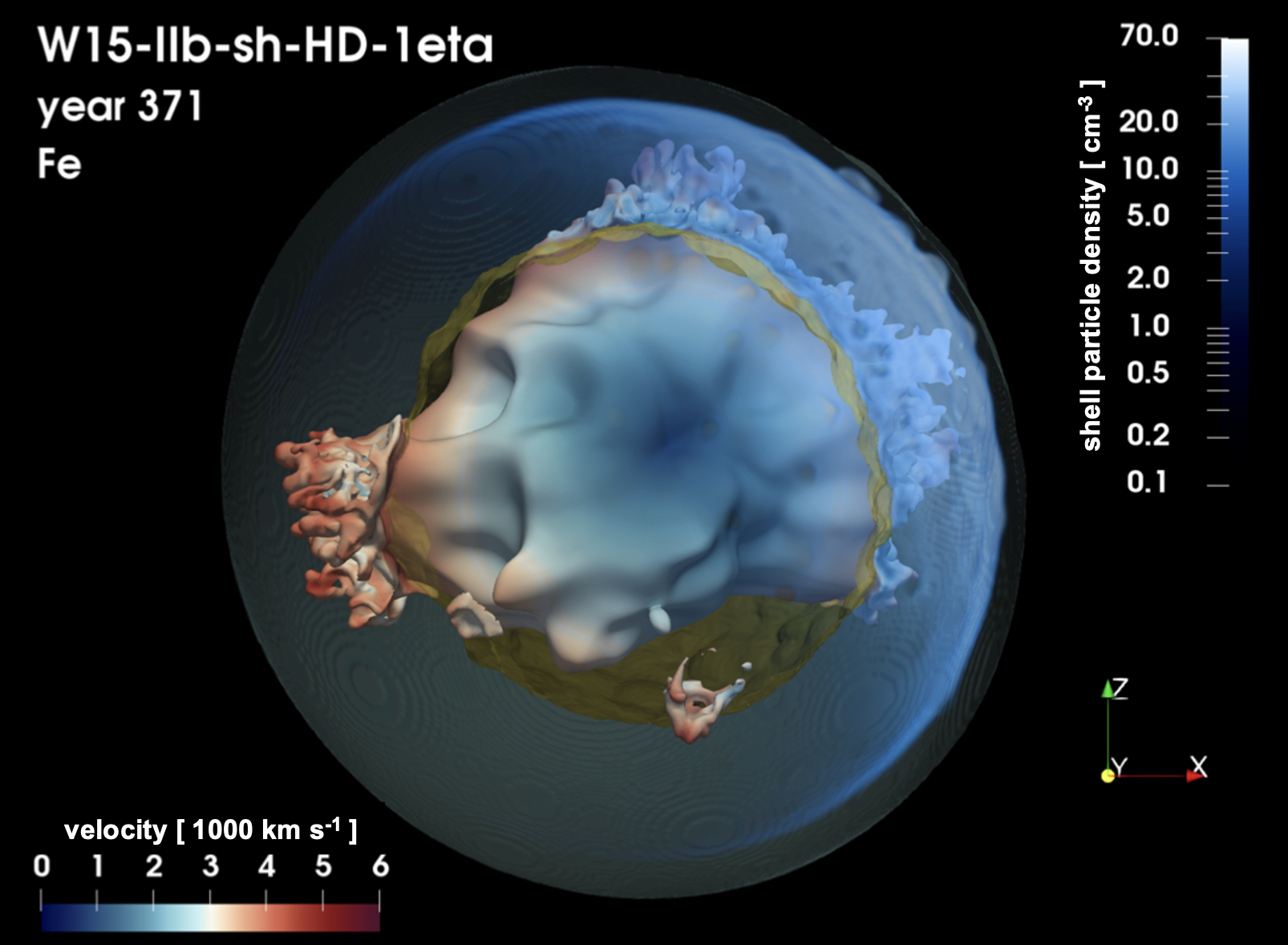}
\caption{Hydrodynamic modeling of the evolution from a neutrino-driven core-collapse SN to the SNR phase at the age of Cas~A. The model describes the interaction of the remnant with an asymmetric circumstellar shell. The figure shows the isosurface of iron distribution (representing a density value at 5\% of the peak density) at the age of Cas~A derived from model W15-IIb-sh-HD-1eta-az in \citealt{2022A&A...666A...2O}. Colors on the isosurface denote radial velocity (with color coding defined at the bottom). Additionally, semi-transparent quasi-spherical surfaces indicate the forward (silver) and reverse (yellow) shocks. The shocked shell is visualized through a volume rendering employing a blue color palette (with color coding on the right); opacity varies proportionally with plasma density.}
\label{fig:model}
\end{figure}

The adopted models also describe the expansion of the blast wave through the wind of the progenitor star (for a detailed description of how the CSM is included, see \citet{2021A&A...645A..66O}). Additionally, model W15-IIb-sh-HD-1eta-az accounts for the interaction of the remnant with an asymmetric circumstellar shell at the age of $\approx 180$~yr, with the densest portion situated in the blueshifted, nearside to the northwest (NW)\cite{2022A&A...666A...2O}. Solar abundances were assumed for the CSM. Figure~\ref{fig:model} shows the distribution of Fe-rich ejecta in the remnant and the interaction with the shell at the age of Cas~A. The remnant is oriented in such a way that the shocked heads of the three extended Fe-rich plumes correspond to the positions of the Fe-rich regions observed in Cas~A. The shell likely resulted from a massive eruption of the progenitor star occurring approximately $10^{4}-10^{5}$ years before collapse. Indeed, this model successfully replicates the large-scale features characterizing the reverse shock observed in Cas~A\cite{2022A&A...666A...2O}, explaining their origin.

The synthesis of {\golem} spectra takes into account several factors (see \citealt{2009A&A...493.1049O, 2019NatAs...3..236M, 2024ApJ...961L...9S} for more details): (i) ejecta abundance derived self-consistently from the SN-SNR simulation; (ii) deviations from equilibrium of ionization and from electron-proton temperature equilibration (see \citealt{2015ApJ...810..168O} for the details of the implementation); (iii) line centroid shift and broadening due to Doppler shift resulting from the bulk motion of plasma along the line of sight; (iv) thermal broadening of emission lines, assuming mass-proportional ion heating (see \citealt{2009A&A...493.1049O} for the details); (v) uniform photoelectric absorption by the interstellar medium, adopting a neutral hydrogen column density value of $N_{H} = 1.5\times 10^{22}$~cm$^{-2}$, appropriate for Cas~A. 

The X-ray spectra were synthesized within the energy range of $[0.1,3]$~keV, employing the non-equilibrium ionization (NEI) emission model \textsc{vnei} from the XSPEC package \cite{1996ASPC..101...17A}, complemented with NEI atomic data sourced from ATOMDB\cite{2001ApJ...556L..91S}. Additionally, we assumed a remnant distance of either $D = 3.4$~kpc, aligning with the distance of Cas~A, or $D = 6$~kpc when simulating the 2000 years old remnant. Subsequently, the synthetic spectra were convolved with the instrumental response of {\golem}, assuming a spectral resolution of 2~eV, resulting in the generation of corresponding focal-plane spectra. Given the close resemblance of the synthetic spectra to actual X-ray spectra, we analysed them using the standard data analysis system employed for true X-ray data (XSPEC v.12.10; \citealt{1996ASPC..101...17A}). It is worth mentioning that the analysis of mock spectra was conducted blind by some team members, who had no prior knowledge of the modeled emitting plasma. This approach was taken to prevent any potential bias in the analysis.
\bigskip

\section{Probing the Supernova Explosion Mechanism with LEM}
\label{sec:analysis}

\begin{figure*}[!th]
\centering
\includegraphics[width=0.97\textwidth, trim={0cm 0cm 0cm 0cm},clip]{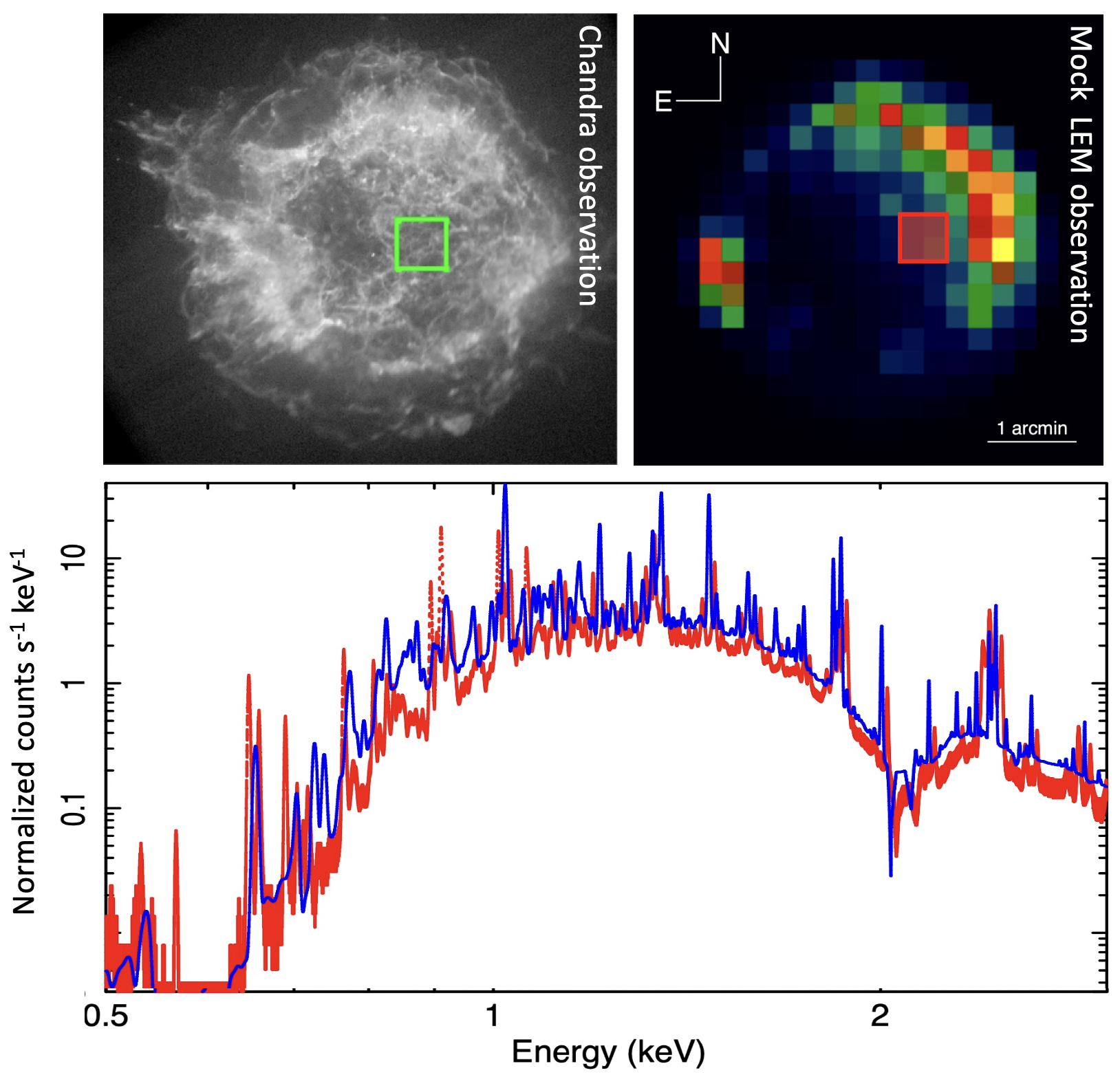}
\caption{Mock {\golem} spectra extracted from a central region in Cas~A. The upper panels illustrate the selected regions with size $40"\times 40"$ for spectrum extraction in both a Chandra image of Cas~A in the [$0.5,8$]~keV band (left) and a synthetic {\golem} image in the [$0.5,3$]~keV band derived from the hydrodynamic model that captures the evolution of Cas~A from the SN\cite{2022A&A...666A...2O} (right). North is up and west is right in both panels. The lower panel presents a comparison between the {\golem} spectrum generated from the best-fit model of Chandra data (blue line) and that synthesized from the hydrodynamic model (red line).}
\label{figure_chandra}
\end{figure*}

When the core of a massive star collapses to a neutron star, an outward shock is generated that propagates into the layers of stellar material experiencing gravitational collapse. Initially, this shock stalls while still deep inside the star. However, the newly formed, hot neutron star radiates a huge flux of neutrinos, a fraction of which heats the postshock matter to thus revive the stalled shock and disrupt the star against its gravitational binding. This neutrino-driven mechanism is supported by non-spherical hydrodynamic instabilities, resulting from convective overturn, global shock oscillations, and turbulent flow fragmentation. These effects initiate an asymmetric explosion, in which the initial ejecta asymmetries seed the growth of secondary mixing phenomena (again caused by hydrodynamic instabilities) at the composition-shell interfaces of the progenitor star. The initial blast-wave asymmetry as well as the secondary radial mixing processes are tightly linked to the structure of the progenitor and are also affected by fundamental characteristics such as stellar rotation and mass loss. For example, the presence of a massive hydrogen envelope may inhibit the expansion of outward reaching plumes of Ni-rich ejecta. Cas~A was the explosion of a progenitor that had stripped nearly all of its hydrogen envelope. It thus offers the unique possibility to closely study the morphology and associated composition of the SN's initial asymmetries that are directly connected to the explosion mechanism.

Hence, the degree of asymmetry in a SN explosion is expected to leave a distinct mark on its remnant. Longslit or integral field unit (IFU) optical spectroscopy of SNe, along with observations of spatially resolved Galactic and Magellanic Cloud remnants, offer means to reconstruct the 3D structure of the SNR ejecta. This provides a direct insight into how energy deposition by the mechanism driving the explosion creates asymmetries in density, velocity, and chemical composition\cite{2015Sci...347..526M, law20, larsson21}, and how and to what extent subsequent radial mixing processes distribute the chemical species asymmetrically through the ejecta. 

The {\golem} microcalorimeter spectrometer is sensitive to emission from shocked SN ejecta, including for instance silicon (Si), oxygen (O), and iron-peak elements. These ejecta originate from various regions of the progenitor star or are synthesized during the explosion, resulting in potentially distinct signatures. Oxygen is the most abundant chemical element ejected from the stellar metal core in core-collapse SNe. Much of it resides at relatively large radii and is typically less affected by the central ``engine'' and asymmetries created at the onset of the explosion. In contrast, most of the expelled silicon is formed by explosive burning of SN-shock-heated oxygen. Also the iron-peak nuclei (ranging from chromium to copper) in the SN ejecta are exclusively formed during the SN explosion, namely by explosive silicon and oxygen burning and by the freeze-out nucleosynthesis in initially extremely hot matter that was directly exposed to the energy input (specifically, neutrino heating) by the SN engine. This innermost iron-group enriched material is predicted to be convectively buoyant and therefore has been observed to be transported outward toward the stellar surface by mixing processes behind the expanding SN shock\cite{2000ApJ...528L.109H, 2021Natur.592..537S}. Consequently, the velocity distribution of the different elements and their spatial asymmetries and inhomogeneities are expected to differ, reflecting the underlying hydrodynamic and nuclear processes during the developing explosion.

In this section, we explore the capability of {\golem} in delivering high-resolution spectra of SNRs and the valuable insights we can glean from analyzing these spectra. Before proceeding with our analysis, however, we first evaluated the realism of the mock {\golem} spectra synthesized from the hydrodynamic simulations. 

More specifically, we investigated the ability of these simulations in replicating the flux and overall spectral characteristics observed in specific regions of Cas A, as captured by Chandra. We identified a central region with size 40\arcsec$\times 40$\arcsec\ within the Chandra observations (upper left panel in Figure~\ref{figure_chandra}), extracted the X-ray spectrum from this area, and subsequently fitted the spectrum using two isothermal \textsc{vnei} components in XSPEC, with ISM absorption assumed at $n_{\rm H} = 1.5\times 10^{22}$~cm$^{-2}$ (appropriate for Cas~A); one of the two isothermal components is assumed to have solar abundances to represent the shocked CSM, while the other allows the abundances of Ne, Mg, Si, S, and Fe to vary freely to account for contributions from the ejecta\footnote{Note that this fit is phenomenological, intended primarily to reproduce the main features of the true spectrum for comparison with the sythetic spectrum from the hydrodynamic simulation rather than provide a rigorous match. Additionally, the results of the spectral fitting are expected to reflect the average properties of the emitting plasma, involving a complex reality made of blending of emission lines, line broadening and Doppler shifts of various gas components in the region analyzed.}. We then convolved the best-fit phenomenological model derived from the Chandra spectrum with the spectral response function of {\golem} \cite{2023HEAD...2011018K} to obtain a Chandra based prediction of the {\golem} data. At the same time, we identified a comparable region within a synthetic {\golem} emission map derived from the hydrodynamic simulation when the remnant is $\approx 350$~years old (upper right panel in the figure) and generated a mock {\golem} spectrum based on the simulation as outlined in Section~\ref{sec:model}. We note that, in this case, we are fully signal-dominated, so adding background emission and instrumental noise does not have a significant impact. Comparison of the two spectra, as shown in the lower panel of Figure~\ref{figure_chandra}, was conducted without any renormalization to align their fluxes. Notably, the hydrodynamic model demonstrated a good reproduction of the flux inferred with Chandra, along with a generally consistent overall spectral shape. Based on these findings and the analysis of the spectrum extracted from the southeast (SE) region shown in Figure~\ref{fig:chandra_SE} (see Section~\ref{sec:casa}), we conclude that the mock {\golem} observations generated using the hydrodynamic model provide a realistic representation of the expected spectra from {\golem} observations of SNRs resembling Cas~A. For our purposes, we synthesized the {\golem} spectra assuming exposure times of either 50~ks or 100~ks.
\bigskip

\subsection{The remnant at the age of Cas~A}
\label{sec:casa}

\begin{figure*}[!th]
\centering
\includegraphics[width=0.99\textwidth, trim={0cm 0cm 0cm 0cm},clip]{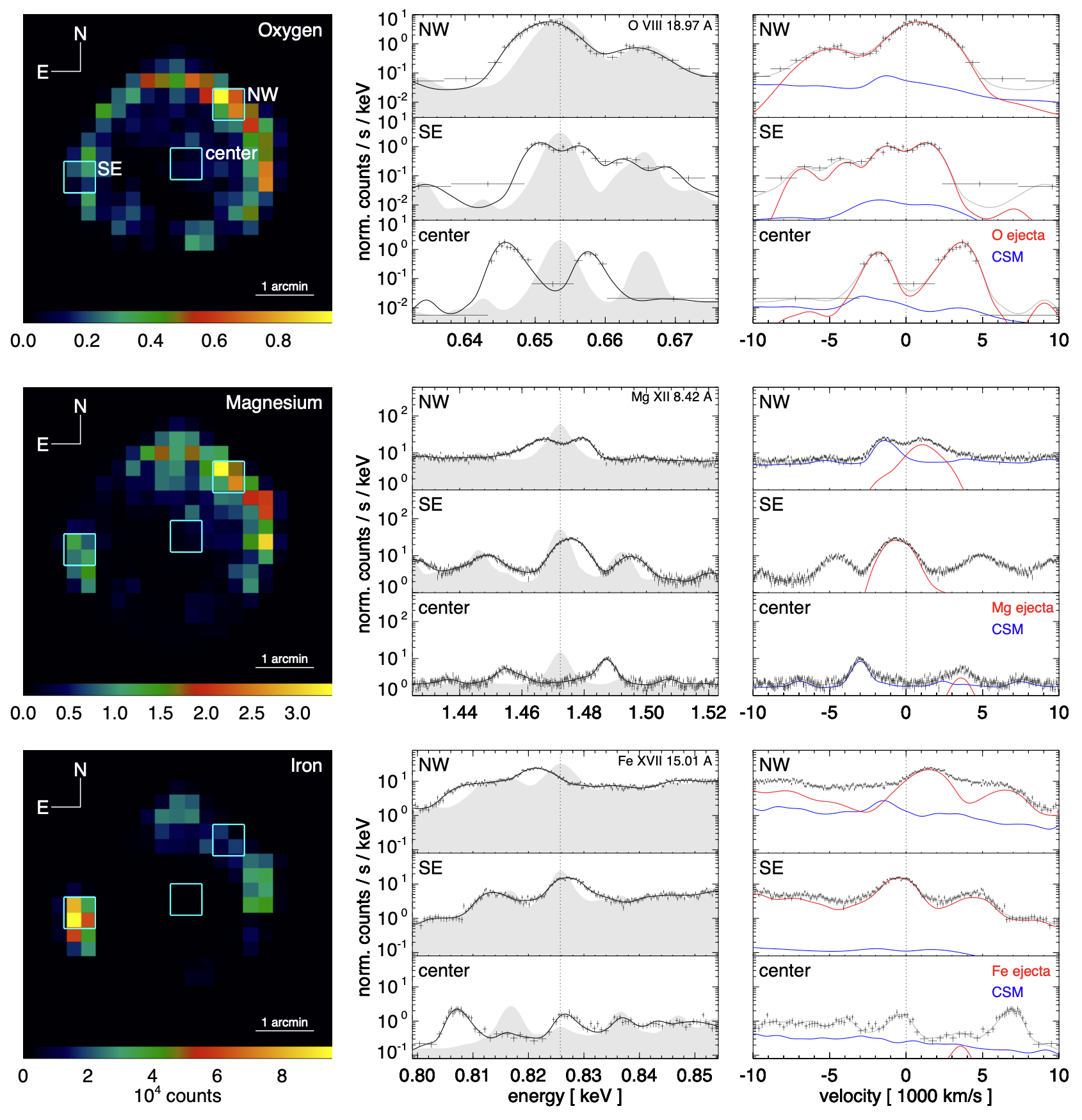}
\caption{Synthetic {\golem} observations of the Cas~A SNR, obtained using the hydrodynamic model W15-IIb-sh-HD-1eta-az in \citealt{2022A&A...666A...2O}. {\em Left column}: Emission maps of Oxygen, Magnesium and Iron (i.e., each map shows the contribution of a specific element to the emission in the {\golem} band). {\em Middle column}: {\golem} spectra centered on specific lines and extracted from the three regions (NW, SE, and centered) evidenced with boxes in the emission maps. The black lines and data points display the spectra with both thermal and Doppler broadening included. In contrast, the shaded area represents the model that includes thermal broadening but does not account for Doppler shifts due to bulk motion. {\em Right column}: Velocity profiles for the lines shown in the middle column, including contributions from CSM (blue lines) and ejecta (red lines) rich in Oxygen (upper panel), Magnesium (middle panel) and Iron (lower panel). The contribution to emission of other ejecta components is omitted for clarity.}
\label{fig:casa}
\end{figure*}

Figure~\ref{fig:casa} shows the mock {\golem} observation of the modeled SNR at the age of Cas~A ($\approx 350$~years) derived from model W15-IIb-sh-HD-1eta-az (namely the model which describes the interaction of the remnant with an asymmetric circumstellar shell). We found that, even in a modest 50~ks observation, {\golem} will be able to map the distribution of elements formed during the progenitor evolution \textit{and} during the explosive nucleosynthesis which occurs during core-collapse. While the differing spatial distribution of these elements (Figure~\ref{fig:casa}, left) reveal, in gross detail, the fact that material such as O and Mg, synthesized during the progenitor evolution, and Fe, the product of explosive Si-burning and/or freeze-out nucleosynthesis, are distributed differently when the star explodes, the real power of {\golem} is shown in the right hand panels of Figure~\ref{fig:casa}. Each panel in the middle column shows spectra (represented by black lines and data points), including Doppler and thermal broadening effects of the lines. These spectra are extracted from selected regions near the NW and SE limbs and in close proximity to the center of the remnant (identified by boxes in the emission maps on the left panels of the figure). The shaded areas in the same panels represent the spectra without considering the kinematics of the emitting medium (i.e., no Doppler effects). The comparison of these areas with the black lines and data points emphasizes the impact of Doppler shifts and line broadening caused by the plasma velocity component along the line of sight, as well as the effects of thermal broadening (included in the shaded areas), on the line profiles. 

The panels in the right column show velocity profiles for the lines presented in the middle column, along with the contribution from the shocked CSM (blue lines) and shocked ejecta (red lines) rich in O (upper panel), Mg (middle panel) and Fe (lower panel). The contribution to emission of other ejecta components is omitted for clarity. Thus, each of these panels illustrates the distribution of ejecta along a specific line of sight, projected into velocity space. The spectra allow the identification of several ejecta components characterized by different shifts and broadening. Additionally, the contribution from the shocked circumstellar shell is evident, as it manifests through a blueshifted component, as observed for instance, in the Mg line extracted from the central region (see the middle row in the figure). In fact, the shell exhibits its highest density on the near side of the remnant, resulting in a prevailing blueshifted line component. Note that, in the middle panel, the contribution of shocked CSM is not visible in the SE region (where the shell is less dense) because below the range of normalized counts~s$^{-1}$~keV$^{-1}$.

\begin{figure*}[!th]
\centering
\includegraphics[width=0.99\textwidth, trim={0cm 0cm 0cm 0cm},clip]{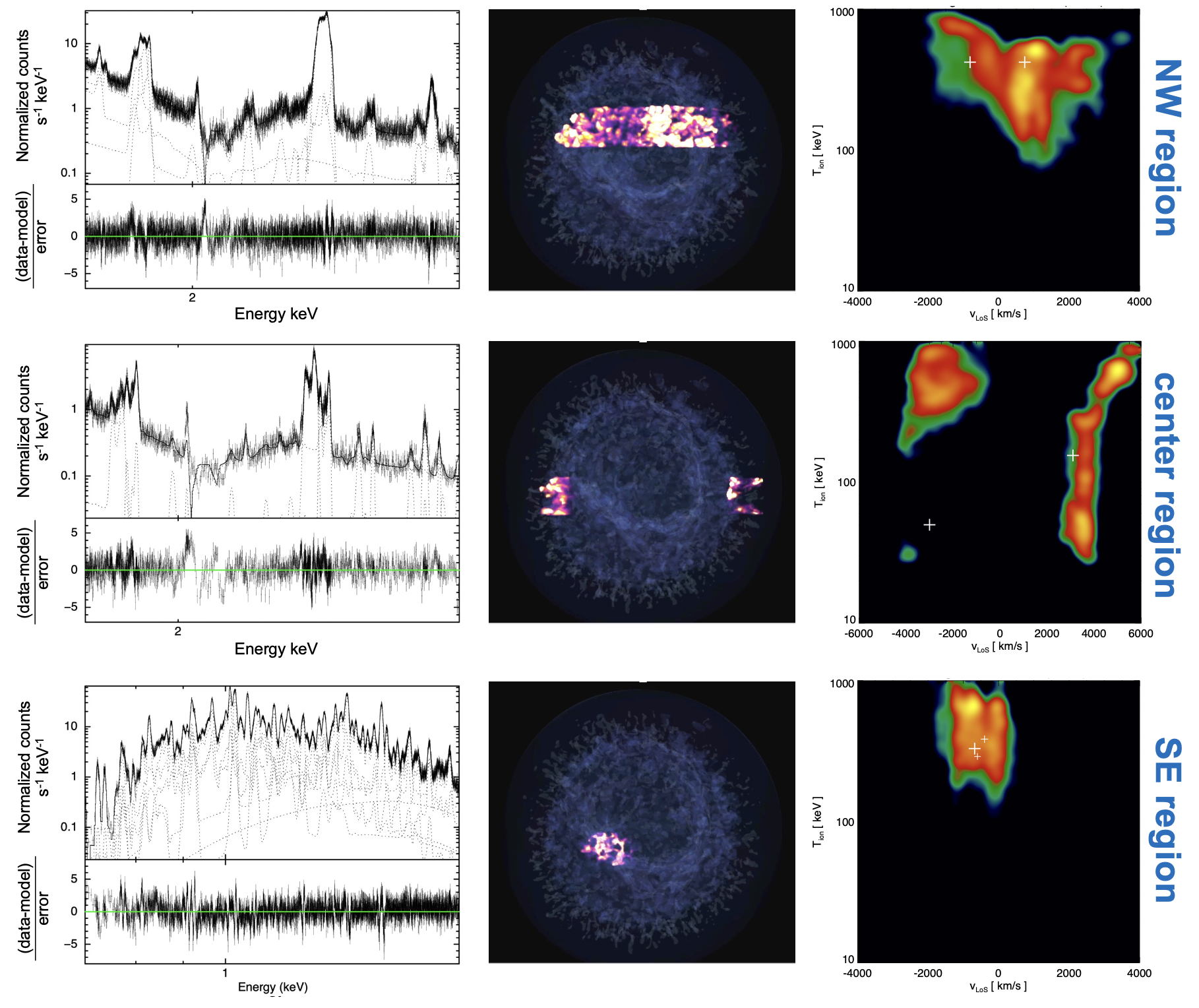}
\caption{Analysis of the synthetic spectra extracted from the regions selected in Figure~\ref{fig:casa}. {\em Left column:} A close-up view of synthetic {\golem} spectra (black symbols) in different energy bands ($[1.7,3]$~keV band in the NW and center regions; $[0.7,1.8]$~keV band in the SE region) to emphasize Si and S lines (NW and center regions) and Fe lines (SE region). The corresponding best-fit models and residuals are also displayed. {\em Center column:} Side view of the ejecta distribution of the remnant (shown in transparent blue; the observer's vantage point is on the left of each panel). The ejecta contributing to the spectra in the selected regions are highlighted with purple-yellow colors. {\em Right column:} Distribution of emission measure versus ion temperature ($kT_{\rm ion}$; Si temperature for the NW and center regions and Fe temperature for the SE region) and velocity component along the line of sight ($v_{\rm los}$) derived from the hydrodynamic model, compared with the results of spectral analysis. The white crosses mark the values inferred from the spectral fitting; the size of the crosses is proportional to the normalization of the corresponding isothermal components.}
\label{fig:spec_analysis}
\end{figure*}

To assess the diagnostic capability of {\golem}, we analyzed the spectra extracted from the three regions shown in Figure~\ref{fig:casa} using XSPEC, treating them as if they were genuine spectra. Here, we discuss the spectra synthesized using a conservative exposure time of 100~ks, instead of the more modest 50~ks. The left column in Figure~\ref{fig:spec_analysis} displays close-up views of the spectra in specific bands that emphasize the profiles of Si, S, and Fe lines. The panels also exhibit the results of spectral fitting with multiple thermal components: the non-equilibrium of ionization optically thin plasma model \textsc{vnei} from XSPEC was used to describe the CSM contribution, and the \textsc{bvrnei} variant (that allows to set the broadening of the lines) was used to characterize the contribution from pure ejecta. Both the spectra extracted from the NW and center regions were fitted in the energy range [$0.7,3$]~keV and required one blueshifted CSM component (\textsc{vnei}) to account for the densest circumstellar shell on the nearside, and two pure-ejecta components (\textsc{bvrnei}). The reduced $\chi^2$ values are 1.3 (4859 d.o.f.) for the NW region and 1.9 (3223 d.o.f.) for the center region. The spectrum extracted from the SE region was fitted in the energy band [$0.7,2$]~keV; it is rich in lines and required two O-Ne-Mg-rich ejecta components, one Si-S-rich ejecta component, and three Fe-rich ejecta components (reduced $\chi^2 =1.6$, 4048 d.o.f.). In this latter case no CSM component was necessary. In fact, the modeled circumstellar shell exhibits only a minor density enhancement in the SE region, making the contribution of shocked CSM to the X-ray spectrum negligible.

The impact of thermal broadening is expected to be more pronounced in the NW region, near the limb, where the bulk motion along the line of sight of the emitting plasma exhibits a reduced range of velocity values. The upper-center panel of Figure~\ref{fig:spec_analysis} highlights the shocked ejecta distributed along the line of sight, as shown in the side-view representation of the ejecta distribution in the NW region, contributing to the spectrum. An almost continuous distribution of ejecta is present with higher densities in the redshifted side, implying that the distribution of velocities along the line of sight is also continuous with the peak slightly redshifted. For this case, the analysis required two pure-ejecta components with velocities $v_1 = -810\pm 30$~km~s$^{-1}$ and $v_2 = +740\pm 30$~km~s$^{-1}$ and ion temperatures $kT_{\rm Si} = 420\pm 10$~keV for Si and $kT_{\rm S} =480\pm 10$~keV for S, the same for the two fitting components. 

We found that the velocity values fit well within the range of values from the hydrodynamic model. The upper-right panel in Fig.~\ref{fig:spec_analysis} compares the values inferred from the spectral fitting with the distribution of emission measure, EM$(kT_{ion}, v_{\rm los})$, versus ion temperature ($kT_{\rm ion}$) and velocity component along the line of sight ($v_{\rm los}$) derived from the hydrodynamic model. The redshifted fitting component aligns neatly with the peak of emission measure at velocities around $\approx 700$~km~s$^{-1}$, whereas the blueshifted component corresponds to a smaller fraction of ejecta with velocities ranging between $-1200$ and $-600$~km~s$^{-1}$. The temperatures inferred from the spectral analysis closely match the emission-measure-weighted values derived from the model, $\overline{kT_{\rm Si}} = 353$~keV and $\overline{kT_{\rm S}} = 403$~keV. This alignment is noteworthy given that the best-fit phenomenological model is a highly approximate representation of the multi-temperature, non-equilibrium emitting plasma. We conclude that, given the conditions observed in the NW region, {\golem} showcases its formidable capability to differentiate between bulk Doppler broadening and thermal broadening of the ion lines. Furthermore, it demonstrates its remarkable power in recovering ion temperatures of shocked ejecta. In fact, the statistical errors on $T_{\rm i}$ are expected to be lower than the uncertainties induced by the stochasticity
in the CSM and ejecta.

The central region is characterized by emission contributions from two large clumps of ejecta, one traveling toward and the other away from the observer with the highest velocity component along the line of sight. These two ejecta clumps are evident in the middle-center panel of Figure~\ref{fig:spec_analysis}. Indeed, the spectral analysis required two ejecta components with high velocities: $v_1 = -3000\pm 100$~km~s$^{-1}$ and $v_2 = +3100\pm 100$~km~s$^{-1}$. These velocities correspond to the two ejecta clumps described by the hydrodynamic model. Notably, the spectral analysis effectively captures the primary velocities of the ejecta clumps, as demonstrated in the middle-right panel of Figure~\ref{fig:spec_analysis}. Once more, our study underscores the capability of {\golem} spectra to accurately retrieve the line-of-sight velocity of shocked ejecta, thereby allowing the 3D reconstruction of the ejecta structure.

At variance with the analysis of the NW region, in the central region, the two fitting components for the ejecta exhibit different values of the ion temperatures: $kT_{\rm Si} = 50\pm 3$~keV and $kT_{\rm S} =57\pm 5$~keV for the blueshifted component, and $kT_{\rm Si} = 155\pm 3$~keV and $kT_{\rm S} =177\pm 5$~keV for the redshifted component. These temperatures do not match the emission-measure-weighted values derived from the hydrodynamic model, $\overline{kT_{\rm Si}} = 370$~keV and $\overline{kT_{\rm S}} = 423$~keV, as also evident from the middle-right panel in the figure showing the EM$(kT_{ion}, v_{\rm los})$ distribution corresponding to the central region. This discrepancy primarily arises from the complex distribution of emission measure, which spans a wide range of temperature values from 50 to 800 keV (particularly evident in the case of the redshifted clump), rendering inadequate a description of the spectrum using only two phenomenological plasma components. Furthermore, the line broadening due to the bulk motion of plasma along the line of sight largely dominates thermal broadening in the central region. Consequently, this complicates the task of constraining the ion temperatures there. In the future, it will be interesting to explore potential methods for enhancing the discriminating power of these fits.

\begin{figure}[!t]
\centering
\includegraphics[width=0.49\textwidth, trim={0cm 0cm 0cm 0cm},clip]{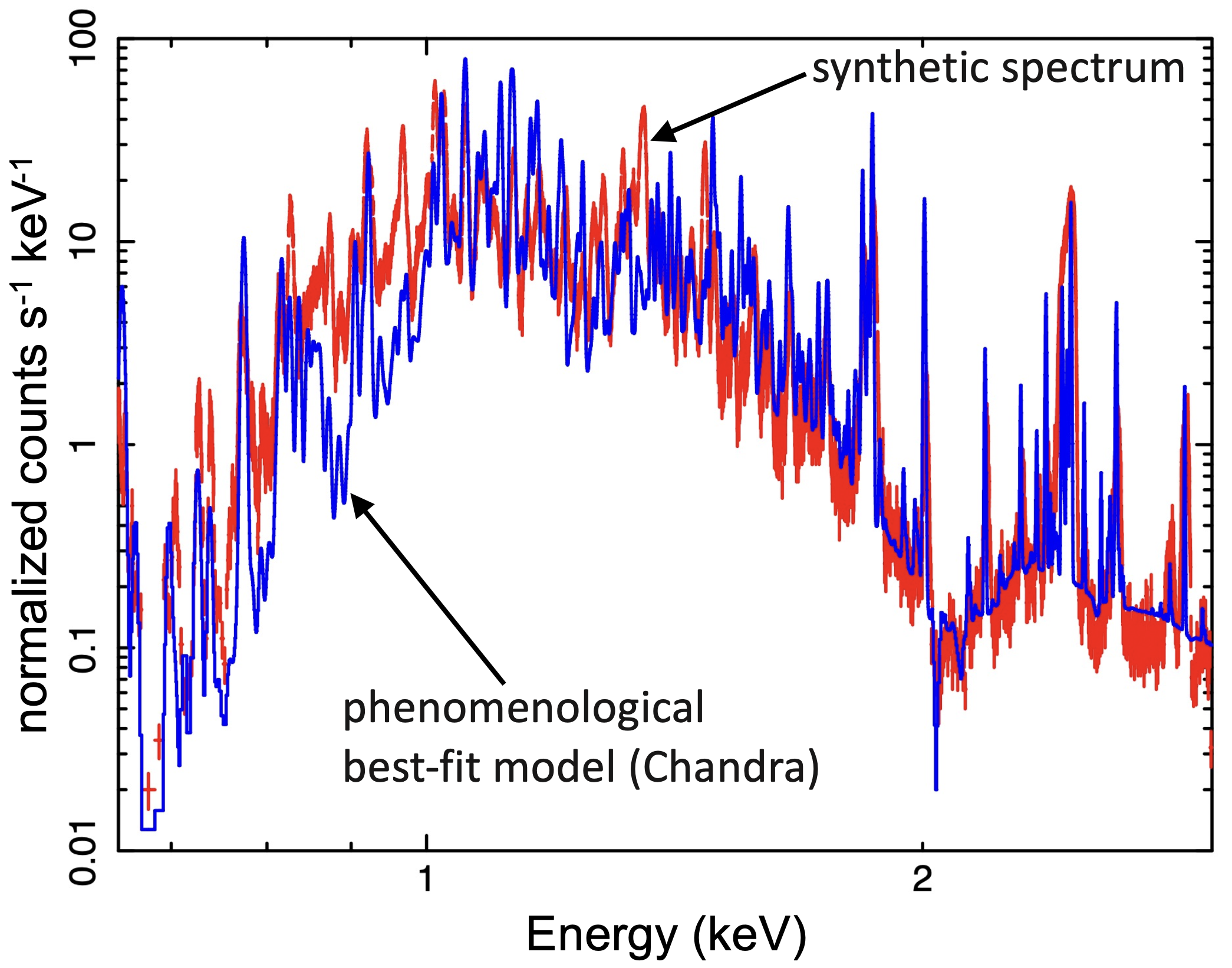}
\caption{Comparison between the {\golem} spectrum synthesized from the best-fit model of a Chandra spectrum (blue line) and that generated from the SNR model (red line), both spectra extracted from a 40\arcsec\ $\times 40$\arcsec\ box centered in the Fe-rich region close to the SE limb.}
\label{fig:chandra_SE}
\end{figure}

The ejecta emission from the SE region originates mainly from an Fe-rich clump with a small velocity component toward the observer (see the lower-center panel of Figure~\ref{fig:spec_analysis}). To assess how accurately the hydrodynamic model represents the conditions within Cas~A, particularly in the Fe-rich region near the SE limb, we first compared the mock {\golem} spectrum extracted from this region with the result of spectral analysis conducted on Chandra observations of Cas~A, similar to the analysis performed for the central region as shown in Figure~\ref{figure_chandra}. More specifically, we selected a region centered within the Fe-rich region near the SE limb in the Chandra observations, performed a spectral fit on the spectrum extracted from this region using isothermal components, and synthesized a corresponding {\golem} spectrum from the best-fit model. The comparison between the {\golem} spectrum synthesized from the best-fit model of the Chandra data and that generated from the SNR model is illustrated in Figure~\ref{fig:chandra_SE}, confirming that the SNR model accurately reproduces the average characteristics of Cas~A spectra extracted from the SE region.

We focused the analysis of the spectrum extracted from the SE region on the $[0.7,2]$~keV spectral band, which is rich in lines. As already mentioned, we fitted the spectrum with multiple \textsc{bvrnei} components: the spectral fitting required one Si-rich ejecta component, two O-Ne-Mg-rich ejecta components, and three Fe-rich ejecta components. We found that the three Fe-rich ejecta components are blueshifted with velocities $v_1 = -600\pm 15$~km\,s$^{-1}$, $v_2 = -400\pm 15$~km\,s$^{-1}$, and $v_3 = -680\pm 50$~km\,s$^{-1}$, consistent with the position of the clump in the nearside of the remnant (see lower-center panel of Figure~\ref{fig:spec_analysis}). The three components are characterized by values of Fe temperatures: $kT_{Fe,1} = 290\pm 15$~keV, $kT_{\rm Fe,2} = 385\pm 15$~keV, and $kT_{\rm Fe,3} = 330\pm 15$~keV. The lower-right panel in Figure~\ref{fig:spec_analysis} compares the results from the spectral fitting with the EM$(kT_{ion}, v_{\rm los})$ distribution derived from the hydrodynamic model. Again, as for the NW region, the match is remarkable in both velocities and temperatures: the spectral analysis has been able to recover the dominant velocities and ion temperatures of the Fe-rich ejecta fragments in the SE region.
\bigskip

\begin{figure*}[!th]
\centering
\includegraphics[width=0.99\textwidth, trim={0cm 0cm 0cm 0cm},clip]{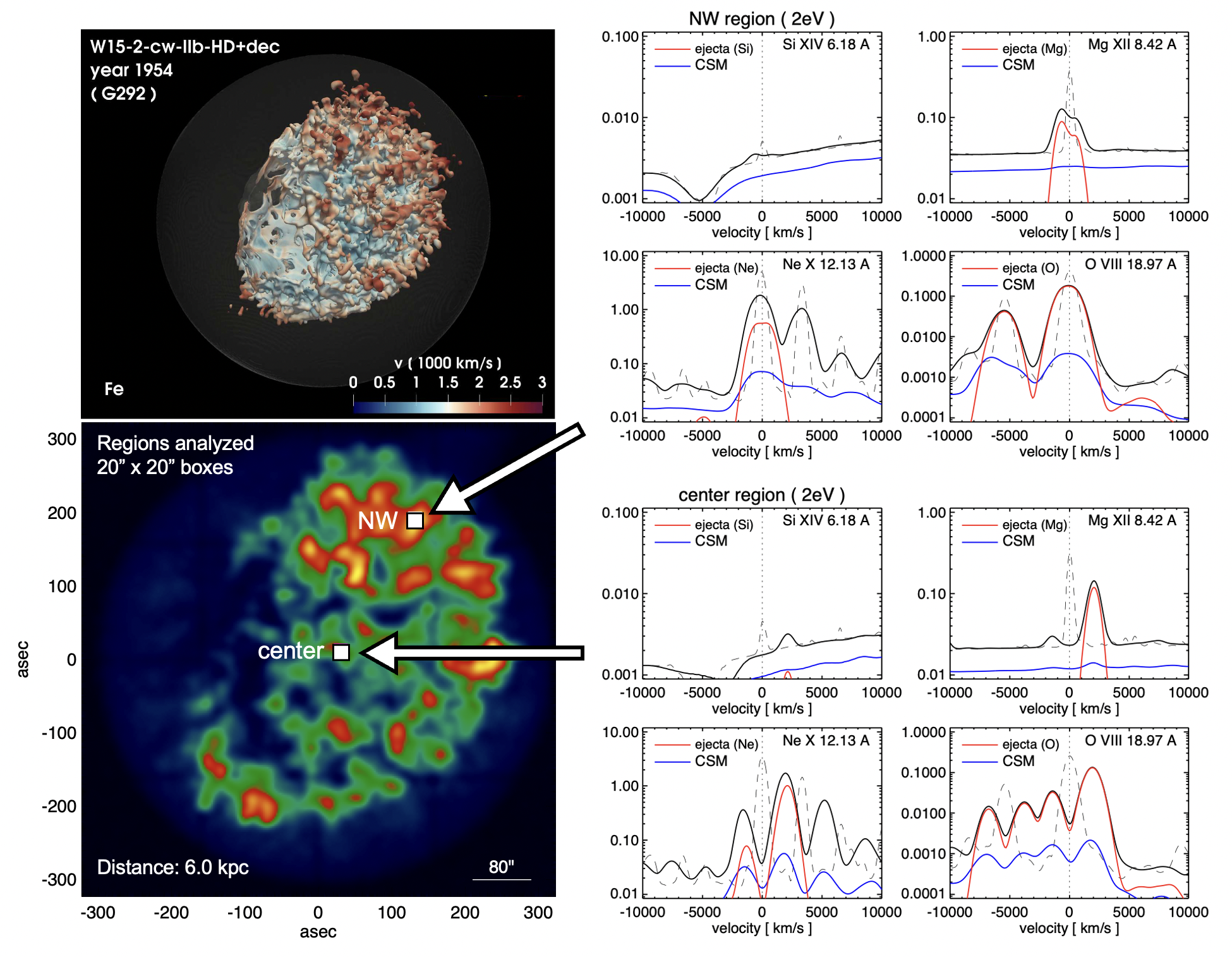}
\caption{Synthetic {\golem} observations of the remnant of a neutrino-driven SN at the age of $\approx 2000$~years derived from model W15-2-cw-IIb-HD+dec in \citealt{2021A&A...645A..66O}. {\em Upper left}: Distribution of Fe-rich ejecta, represented by an isosurface corresponding to a value of Fe density at 5\% of the peak density; the colors indicate radial velocity in units of 1000~km~s$^{-1}$ on the isosurface, with the color code defined at the bottom of the panel. The semi-transparent quasi-spherical surface represents the forward shock. {\em Lower left}: Emission map in the [$0.5, 3$]~keV {\golem} bandwidth, assuming a spatial resolution of 15\arcsec\ and a remnant distance of 6~kpc. {\em Right panels}: Synthetic {\golem} velocity profiles for the lines of Silicon, Magnesium, Neon, and Oxygen centered on the H-like resonance lines of Si XIV, Mg XII, Ne X, and O VIII (solid black lines). These profiles are extracted from the regions highlighted with boxes in the emission map (lower left panel). The panels also show the contributions of shocked ejecta material (red lines) and shocked CSM (blue lines). The dashed black lines show the synthetic spectra that include thermal broadening but do not account for Doppler shifts dur to bulk motion.}
\label{snr_2000yr}
\end{figure*}

\subsection{The Remnant at the Age of 2000 Years}

The hydrodynamic models were also used to explore the {\golem} potential in investigating a more mature SNR compared to Cas~A. To do this, we examined model W15-2-cw-IIb-HD+dec from \citealt{2021A&A...645A..66O} at an age of approximately 2000 years. In this model, the remnant is assumed to expand through a $r^{-2}$ wind originating from the progenitor star before its collapse, and no dense circumstellar shell is included, as assumed for Cas~A (refer to Section~\ref{sec:casa}). The upper-left panel of Figure~\ref{snr_2000yr} shows the modeled distribution of Fe-rich ejecta derived from this model, assuming a different arbitrary orientation for the three Fe-rich plumes than the one adopted for Cas~A in Section~\ref{sec:casa}. 

The mock spectra (right panels in the figure) are synthesized assuming a remnant distance of 6~kpc, comparable to that of the Galactic core-collapse SNR G292.0+1.8\cite{2003ApJ...594..326G} (hereafter referred to as G292). Furthermore, the remnant was oriented in space to roughly resemble the observed distribution of ejecta in G292 (see \citealt{2021A&A...645A..66O} for more details). In fact, G292 shares many similarities with Cas~A as an oxygen-rich SNR but is more evolved, with an estimated age of $\gtrsim 2000$ years\cite{2022ApJ...932..117L}, a timeframe comparable to the hydrodynamic simulation used here. Unlike Cas~A, G292 contains an energetic pulsar with a high imparted kick velocity\cite{2022ApJ...932..117L}, and is thought to be the remnant of a highly stripped progenitor\cite{temim22}. Given these differences, it is important to express some words of caution about this model. In fact, the model assumes a progenitor stripped star with a specific He-core mass and a SN explosion energy, both tailored to approximate Cas~A and it is not expected to be applicable to G292. Nevertheless, the close alignment between the average forward and reverse shock radii derived from the simulation and those inferred for G292 from observations renders this case particularly interesting for testing the capabilities of {\golem} in the case of a remnant with age and other characteristics similar to those of G292.

The synthesis of {\golem} spectra follows again the methodology outlined in Section~\ref{sec:model}. We focused on two distinct regions, each spanning 20\arcsec$\times 20$\arcsec, for extracting the {\golem} spectra (as shown in the lower-left panel of Figure~\ref{snr_2000yr}): one located near the NW limb and the other near the center of the remnant. In this case we did not fit the extracted spectra as described in Section~\ref{sec:casa}, but instead focused on examining the shifts and broadening of the emission lines. 

We found that the spectral lines appear to be narrower compared to those of the younger Cas~A-like remnant, primarily due to the deceleration of the ejecta that occurred during the remnant expansion, resulting in a reduced spread of their velocities along the line of sight. In both selected regions, there is negligible contribution of shocked ejecta to the Si\,XIV line, while the ejecta make a significant contribution to lines of Mg\,XII, Ne\,X, and O\,VIII. 

In the NW region, lines originating from the ejecta exhibit moderate broadening and no significant shift (although the Mg\,XII line suggests a plasma component blueshifted at velocities around $-1000$~km~s$^{-1}$), indicating that ejecta moving both away from and towards the observer are uniformly distributed along the line of sight. This result is expected because, in regions close to the limb, the component of ejecta velocity projected along the line of sight tends to be lower compared to that in regions near the center of the remnant. Consequently, the ejecta are expected to be distributed nearly continuously from the blueshifted to the redshifted side of the remnant, as in the case observed in the NW region of Cas~A (refer to the upper panels of Figure \ref{fig:spec_analysis}).

The highest bulk velocities of the ejecta are observed in the O\,VIII lines, with their wings extending up to velocities of approximately 3000~km~s$^{-1}$. There is no notable presence of shocked CSM observed in the Si\,XIV and Mg\,XII lines, while a discernible contribution is detected in the other two lines. In these cases, the lines originating from shocked CSM exhibit a slightly broader profile compared to those from ejecta, with their wings extending to velocities of approximately 4000~km~s$^{-1}$. This higher line broadening is likely due to the accelerated motion of the shocked CSM by the faster outer layers of the ejecta.

In the central region, the spectral lines exhibit narrower profiles compared to those in the NW region and are characterized by notable Doppler shifts. Specifically, the Mg\,XII line displays a single component redshifted to velocities of $\sim 2000$~km~s$^{-1}$, while the Ne\,X and O\,VIII lines show two components: one redshifted to velocities of $2000$~km~s$^{-1}$ and another, with a lower flux, blueshifted to velocities of $-1500$~km~s$^{-1}$ (refer to the lower-right panels in Figure \ref{snr_2000yr}). Similar to the case observed in Cas~A, the two line components in the spectrum extracted from the central region correspond to two distinct ejecta clumps, with one moving away from and the other toward the observer. The contribution of shocked CSM to the emission lines is also observed with similar shifts but lower fluxes, suggesting that the shocked CSM possesses velocities along the line of sight comparable to those of the shocked ejecta.
\bigskip

\section{Summary and Conclusions}
\label{sec:conclusions}

We investigated the potential of {\golem} to collect high-resolution X-ray spectra of spatially resolved SNRs, aiming to evaluate the information that will be possible to extract from spectral analysis, in particular for the study of core-collapse SN physics. To accomplish this, we exploited a state-of-the-art 3D hydrodynamic model, describing the evolution of a neutrino-driven SN from core-collapse to the formation of a mature remnant. Using this model, we synthesized realistic mock spectra representative of those {\golem} would capture during an observation. These mock spectra account for various factors, such as Doppler shift and broadening of emission lines due to the bulk motion of the emitting plasma along the line of sight, broadening of lines due to the thermal motion of ions, deviations from ionization equilibrium and electron-proton temperature equilibrium, and photoelectric absorption by the interstellar medium.

The mock spectra were synthesized to represent two distinct evolutionary phases of a SNR: at approximately 350 years and 2000 years post-explosion. In the former scenario, the hydrodynamic model generates a remnant that quite well replicates some of the large-scale properties of the Cas~A SNR. In the latter scenario, the model simulates a mature oxygen-rich SNR with age and other attributes similar to those observed in the G292 SNR. Consequently, the mock spectra generated in these instances offer a representative glimpse into the spectra that could potentially be obtained with {\golem}. 

We analyzed the spectra as if they were genuine spectra. Using phenomenological multiple thermal components in XSPEC, we fitted the spectra accurately. Our investigation focused particularly on SNRs resembling oxygen-rich SNRs as Cas~A or G292, revealing the remarkable potential of {\golem}. Our findings suggest that {\golem} will be able to discern emission lines stemming from a variety of chemical species with high precision (see Figures~\ref{fig:casa} and \ref{snr_2000yr}). Moreover, {\golem} demonstrates an ability to distinguish between line components exhibiting diverse Doppler shifts and broadening, offering unprecedented insights into the velocity distribution of the emitting shocked plasma along the line of sight (see Figures~\ref{fig:spec_analysis} and \ref{snr_2000yr}). This capability opens avenues for performing 3D kinematic reconstruction of the X-ray-emitting, metal-rich ejecta, similar to methodologies employed in optical band studies (e.g., see for instance \citealt{2013ApJ...772..134M, law20}). Hence, this groundbreaking resolution empowers researchers with the opportunity to reconstruct the intricate 3D structure of shocked ejecta, thereby unlocking a wealth of information crucial for advancing our understanding of the mechanisms of core-collapse SNe, in particular neutrino-driven vs.\ jet-driven, and their nucleosynthetic and morphological fingerprints. 

The spectral analysis has also revealed that, in the case of an interaction between the remnant and a dense and inhomogeneous CSM, {\golem} demonstrates its potential to offer insights into the structure and density distribution of the CSM. An illustrative example lies in the modeling of the interaction between Cas~A and an asymmetric circumstellar shell\cite{2022A&A...666A...2O}, where synthetic {\golem} spectra clearly reveal an emitting plasma component characterized by solar abundance (see Figure~\ref{fig:casa}). Notably, this component exhibits a blueshift, particularly prominent at the center and NW limb of the remnant, consistent with the SNR model's prediction attributing the densest portion of the shell to the nearside NW quadrant. This evidence paves the way for reconstructing the CSM sculpted by the progenitor star's winds in the centuries to millennia preceding the SN event. Such information holds pivotal importance in constraining the mass-loss history of the progenitor star during its final lifecycle stages, offering crucial insights into the elusive mass-loss mechanisms that characterize the concluding phase of evolution for massive stars.

It is worth emphasizing the broad applicability of the diagnostic capabilities offered by {\golem} to various remnants of core-collapse SNe. For spatially extended SNRs like Puppis~A, the use of {\golem}'s high-resolution spectroscopy will enable researchers to analyze the line profiles and Doppler shifts, allowing for a comprehensive characterization of the shock structure and dynamics within these remnants. This analysis potentially yields critical parameters such as shock velocities, temperature gradients, and density variations, all pivotal for deciphering the intricate physics governing SN explosions. Moreover, for remnants interacting with dense and irregular environments like IC~443 and VRO 42.05.01, the high-resolution spectra obtained with {\golem} can facilitate the identification and characterization of these interactions. This can be important to shed light on the structure and geometry of the ambient environment but also to highlight the impact of the environment on the morphology and evolution of the remnants, for example, affecting the shock dynamics and leading to mixed morphology features \cite{2021A&A...649A..14U, 2021A&A...654A.167U}. Furthermore, high-resolution spectra can reveal signatures of molecular and atomic gas associated with the remnants, offering insights into the pre-SN environment and the effects of the explosion on the surrounding medium. This encompasses a detailed study of the kinematics and distribution of gas components, enriching our understanding of the dynamic interplay between the remnants and their surroundings.

Our investigation has revealed another important capability of {\golem} spectra: the possibility to discriminate between bulk Doppler broadening and thermal broadening of ion lines. Such discrimination becomes feasible near the limb of the remnant, where Doppler effects exhibit less prominence owing to a narrower spread in the velocity distributions along the line of sight and, therefore, to a reduced broadening. Exploiting this advantage, spectral fitting enabled us to tightly constrain ion temperatures with remarkable precision, thereby recovering emission-measure-weighted average temperatures of selected ions. Consequently, we anticipate that the analysis of {\golem} spectra will allow the investigation of ion heating processes at the shock fronts of SNRs, as well as the subsequent cooling of these ions in the post-shock region. This capability promises to deepen our understanding of the intricate dynamics governing astrophysical shocks.

\bigskip
\noindent
{\bf Acknowledgments}\\
The \textsc{pluto} code is developed at the Turin Astronomical Observatory (Italy) in collaboration with the Department of General Physics of Turin University (Italy) and the SCAI Department of CINECA (Italy).
S.O., M.M., and F.B. acknowledge financial contribution from the PRIN 2022 (20224MNC5A) - ``Life, death and after-death of massive stars'' funded by European Union – Next Generation EU, and the INAF Theory Grant ``Supernova remnants as probes for the structure and mass-loss history of the progenitor systems''.

\small
\vspace{-6mm}
\parindent=0cm
\baselineskip=12pt

\bibliography{references}{}
\bibliographystyle{aasjournal}

\end{document}